\documentclass[onefignum,onetabnum]{siamart171218}

\newcommand\m[1]{\begin{pmatrix}#1\end{pmatrix}} 

\usepackage{breakcites}
\newcommand\one{\overrightarrow{1}}

\ifpdf
\hypersetup{
  pdftitle={Schur Complementary Allocation: A Unification of Hierarchical Risk Parity and Minimum Variance Portfolios},
  pdfauthor={Peter Cotton}
}
\fi

\title{Schur Complementary Allocation: \\
 A Unification of Hierarchical Risk Parity and \\
 Minimum Variance Portfolios}

\author{Peter Cotton}

\begin{document}
\maketitle

\begin{quote}
Despite many attempts to make optimization-based portfolio construction in the spirit of  \cite{Markowitz1952PortfolioFinance} robust and approachable, it is far from universally adopted. Meanwhile, the collection of more heuristic divide-and-conquer approaches was revitalized by  \cite{DePrado2016BuildingSample} where Hierarchical Risk Parity (HRP) was introduced. This paper reveals the hidden connection between these seemingly disparate approaches.  
\end{quote}

%

\section{Introduction}

The practice of portfolio construction might be said to split into two schools according to whether optimization is cautiously employed or studiously avoided. One school builds upon \cite{Markowitz1952PortfolioFinance} where optimization was first suggested, and this body of ideas has come to be called Modern Portfolio Theory (MPT). It is by no means universally used, however, and in practice the creation of portfolios today often bears more affinity to advice predating Markowitz. Benjamin Graham in {\em The Intelligent Investor} advised clients to rebalance between stocks and bonds periodically and within each, take steps to avoid undue concentration \cite{graham1949intelligent}. As can be seen even in this example, optimization-free construction of diversified portfolios often takes on a hierarchical, or recursive nature.  

Preference for schemes eschewing optimization persists well after  \cite{Markowitz1952PortfolioFinance} and well after convenient, free, scalable software for the same has been broadly disseminated. This puzzle was considered by Richard O. Michaud  \cite{michaud1989markowitz} thirty years ago and he termed it ``enigma'' - surely even more so now. Michaud's view was that ``indifference of investment practitioners to mean-variance optimization technology, despite its theoretical appeal, is understandable in many cases'' - in part due to a tendency towards ``error maximization'' already studied by Jobson and Korkie \cite{jobson1980estimation} a decade prior. A substantial literature devoted to shrinkage, constraints and other counter-measures has evolved (see \cite{markov2023portfolio} for a recent example employing regularization) but has not yet won the day. The complexity also adds to what Michaud termed the ``conceptual demands placed on managers'' by optimization. 

Moreover practitioners and their clients often seek financial meaning in portfolios and do not always find it in the output of a potentially brittle optimization. It would be unsurprising if the second, optimization-eschewing school of thought were to remain the more popular even in the absence of new ideas to invigorate it. In fact, at least one important idea has helped that cause and in  \cite{DePrado2016BuildingSample} the method of Hierarchical Risk Parity (HRP) was introduced. This top-down method makes only defensive, somewhat indirect use of covariance information via ``quasi-diagonalization'': which is to say a reordering of assets. Permuting indices of the covariance matrix can be viewed as attempt to diagonalize using only permutation matrices, and this hints at, but does not provide, a clean connection to optimization. 

This paper constitutes, we hope, a surprising twist in this battle of ideas. For here we reveal the mathematical connection between the optimization school and the hierarchical. To do that we provide a top-down allocation scheme capable of replicating the minimum variance portfolio. The crucial idea is that covariance matrices can be modified as we proceed downward - in a manner that finds a clean financial and probabilistic interpretation. A new class of hierarchical portfolio methodologies is thereby presented that can borrow the strengths of \cite{DePrado2016BuildingSample} and other top-down technique, as required, but at the same time make a literal connection back to optimization. 

The value of the continuum described herein can be appreciated by the following simplified thought experiment that those in both schools can agree on, and that those in other fields dealing with isomorphic problems (such as minimum variance model combination) might also appreciate. For imagine we zero out excess stock returns and care only about arriving at a minimum variance portfolio. We vary the quality of the information we have, that is, our ability to discern the true covariance matrix, and note that in the limit of perfect information, we best use explicit optimization. But in the other limit, where deviations in our covariance estimates from some grand average are entirely self-deception, we would be better off employing Hierarchical Risk Parity (or some other diversification heuristic, even uniform weights).

Clearly, neither school dominates the other. They each have their territory. 

\section{The hierarchical school}
\label{sec:literature}
A comprehensive survey of optimization-based portfolio construction is largely beyond our scope, and here we focus on the the more niche hierarchical literature. To touch on the former briefly, our informal explorations utilized a growing collection of covariance estimators provided in \cite{cottonprecise}. These representing a superset of shrinkage and other covariance methods in Scikit-Learn \cite{scikit-learn} (standard approaches include Graphical Lasso \cite{friedman2008sparse} and shrinkage per \cite{ledoit2003improved}). These are combined with optimization-based portfolio methods from PyPortfolioOpt \cite{pyportfoliopt} and RiskFolioLib \cite{egozcue2020riskfoliolib}, two popular Python portfolio optimization packages. 

Nor shall we survey applications for minimizing the variance of a linear combination of random variables whose joint distribution is uncertain - though this is obviously quite general and an old problem in forecasting (\cite{bunn2001statistical}), linear ensembles of machine learning models, and elsewhere. In \cite{DePrado2016BuildingSample} it was furthermore remarked that the core idea of HRP might be useful in other problem domains where inversion of a covariance matrix is problematic or impossible. That even more general assertion, if true, almost certainly applies to the approach we introduce in Section \ref{sec:schur} due to the hierarchical nature of the algorithm and its proximity to HRP.    

To describe the literature for hierarchical allocation, we lay out some minimal notation. We limit ourselves to bisect and conquer methods for assigning weights to assets, coinciding with \cite{DePrado2016BuildingSample} where a unit of wealth to be invested is first split among two subgroups of assets using an inter-group allocation ratio $1/\nu(A): 1/\nu(D)$. We follow this inverse convention since $\nu(A)$ and $\nu(D)$ will find interpretation as a portfolio variance, or another measure of inverse investment fitness.  

In that paper, top-down allocation schemes were described in procedural fashion but here we prefer a terse functional expression where intra-group allocations $w(A)$ and $w(D)$ are determined by a method $w()$ that calls out to a terminating portfolio algorithm $w_{term}()$ once the dimension is sufficiently small. We write  
\begin{equation}
\label{eqn:topdownbasic}
w(\Sigma) \propto \left\{  \begin{array}{ll} \m{ 
                      \frac{1}{\nu(A)} w(A)  \\
                       \frac{1}{\nu(D)} w(D)
                   }&, {\rm if}\ \dim(\Sigma)>m  \\
                   
                   w_{term}(\Sigma) &,  {\rm otherwise}
                  \end{array} \right.
\end{equation}
and it is implicit that the right hand side will be normalized $\sum_i w_i =1$. Here we subdivide {\em after a reordering of assets} that places similar securities close to one another. This seriation step, denoted $\pi$ is implicit, which is to say that in \ref{eqn:topdownbasic} the matrix $\Sigma$ uses permuted coordinates without loss of generality.\footnote{One applies the permutation to the covariance matrix $\Sigma$, determines $w$, and then applies the inverse permutation to solve the original problem.} The notation $A$ and $D$ refers to submatrices of $\Sigma$: 
\begin{equation}
\label{eqn:split}
  \Sigma = \m{ A & B  \\
               C=B^T & D }
\end{equation}
This splitting can occur multiple times as \ref{eqn:topdownbasic} is called recursively.  For avoidance of doubt, in \ref{eqn:topdownbasic} both $w(A)$ and $w(D)$ are column vectors and $\nu(A)$ and $\nu(D)$ are scalar. 

Given these components, top-down portfolio allocation can evidently take on many flavors depending on the choice of inverse fitness function $\nu$, terminating portfolio method $w_{term}$, and the seriation $\pi$. The method of estimating $\Sigma$ is not central to this work, although opinions are contained in \cite{LopezdePrado2019EstimationMatrices} for example, and also in \cite{Molyboga2020AManagement} where Ledoit-Wolf shrinkage is advocated specifically in the context of HRP.

Other examples of hierarchical allocation include  \cite{Raffinot2016HierarchicalAllocation} and 
\cite{Raffinot2018ThePortfolio}. Here deviations from \cite{DePrado2016BuildingSample} occur in the the  modifications to the measure of fitness $\nu$; choice of recursion termination criteria; and changes to the terminal choice of portfolio $w_{term}$. Work has generally supported the second school of thought, as we have termed it, as with \cite{Rane2022FinancialAnalysis} but \cite{Kaczmarek2022BuildingPredictions} is an exception, since this paper pushes back on the empirical superiority of hierarchical risk parity over optimization. 

The seriation proposed in \cite{DePrado2016BuildingSample} depends on the (dis)similarity metric. This choice is explored by \cite{Lohre2020HierarchicalAllocations} although in other respects this work follows the established pattern. So too \cite{Sen2021HierarchicalStocks} and \cite{Nourahmadi2021HierarchicalExchange} who appear to agree with Lopez de Prado's sentiment regarding out of sample performance. There is a potential combinatorial explosion in this area as are many ways of performing clustering and partitioning of a portfolio, many approaches to estimating correlations or covariances (e.g. \cite{SunPalomar2023}), and a diversity of measures of similarity that feed the seriation. See \cite{Marti2021AMarkets} for a survey and \cite{Liiv2010ReviewOverview} for an historical review of seriation in statistics.   

Other changes to hierarchical allocation are less ``plug and play'' and include \cite{Pfitzinger2021AAllocation} where the similarity of assets near the bisection boundaries is considered, and it is claimed that one can reduce the loss of information during this process by remaining conscious of unfortunate break points. Another more invasive change is the introduction of constraints in \cite{pfitzinger2019constrained}.

\section{Critique of Hierarchical Risk Parity}
\label{sec:critique}

One can contrast HRP to the theoretical minimum variance long-short portfolio whose weights sum to unity. The solution before normalization is well known:
\begin{equation}
\label{eqn:one}
    w \propto \Sigma^{-1} \one
\end{equation}
where $\one$ is a vector of ones. The approach \ref{eqn:topdownbasic} avoids inversion of the covariance matrix but a different price is to be paid: the information in $B$ is largely thrown away. The novelty of \cite{DePrado2016BuildingSample} lies in the defensive use of $\Sigma$ as a means of generating a distance metric between assets and, subsequently, a reordering that reduces the typical size of entries in the $B$ matrix. This device helps but it does not, unfortunately, fully eliminate this discarding of information. So, one obvious challenge with top down allocation is that it can leave a stock in the first group of assets that is reasonably strongly correlated with several in the second group (the seriation notwithstanding). 

Another critique of hierarchical allocation is that it violates symmetry. Take
$$
\Sigma = \m{1 & \rho & \rho \\ \rho & 1 & \rho \\ \rho & \rho & 1}
$$
where, if we also also assume symmetry in asset returns, it is obvious that wealth should be should be evenly allocated. However if we apply a bisection approach then we must break the symmetry. As shown in Section \ref{sec:examples} we may discover that hierarchical allocation eventually leads to a portfolio:
$$
  w = \frac{1}{3+\rho} \m{ 1 \\ 1 \\ 1 + \rho }
$$
which is unacceptable if $\rho \ne 0$. A possible solution uses only the diagonal sub-covariance entries, such as with
$$
   \nu(\Sigma) = \sum_i \Sigma_{ii}^2
$$
On the other hand a diagonal allocation rule such as this will fail the next elementary example:
$$
\Sigma = \m{1 & \rho & 0 \\ \rho & 1 & 0 \\ 0 & 0 & 1}
$$
for the opposite reason: the allocation will ignore the off-diagonal terms and return a symmetric portfolio $w=\m{1/3 & 1/3 & 1/3}^T$. More generally speaking, diagonal allocation will over-allocate to the sub-portfolios with highest internal correlation, and attract dollars where diversification is in part illusory.

\section{Hierarchical Minimum Variance (HMV)}
\label{sec:schur}

In order to address this inelegance a new type of top-down allocation methodology is now introduced. It departs from the work mentioned in Section \ref{sec:literature} not by choice of seriation, terminal portfolio or sub-portfolio fitness function $\nu$, but rather by breaking an assumption so innucuous it has gone unstated in the literature. All prior work to our knowledge has tacitly assumed that the original $A$ and $D$ sub-covariance matrices should be passed to the recursive step unchanged although this does, in fact, warrant financial and mathematical scrutiny. 

Using the same way of representing hierarchical allocation as in \ref{eqn:topdownbasic}, we now introduce a new family of top-down methods with the following change:
\begin{equation}
\label{eqn:topdownschur}
w(\Sigma) \propto \m{ 
            \frac{1}{\nu\left(A' \right)} w\left(A'' \right) \\
            \frac{1}{\nu\left(D' \right)} w\left(D'' \right) \\
          }                            
\end{equation}
where, in contrast with \ref{eqn:topdownbasic}, the $A'=A'(\Sigma)$ and $A''=A''(\Sigma)$ are modifications of $A$ designed to incorporate some information from the full covariance matrix $\Sigma$. Likewise $D$ is also augmented in two different ways. 

A practical remark for software developers and practitioners is made here. The new matrices $A'$, $A''$, $D'$ and $D''$ are straightforward matrix functions of $A$, $B$ and $D$ comprising the original matrix $\Sigma$, so this can be a relatively non-invasive change to any existing implementation of, say, Hierarchical Risk Parity (see \cite{cottonprecise} for examples). The sub-covariance matrix $A$ is swapped out for an augmented matrix $A''$: 
\begin{equation}
\label{eqn:adashdash}
  A'' = \frac{ A - BD^{-1}C }{b_A b_A^T}
\end{equation}
where pointwise division is indicated, and where
\begin{equation}
\label{eqn:ba}
b_A = \one - AC^{-1} \one.
\end{equation}
Similarly matrix $A'$ is the result of ``pointwise multiplication of the Schur complement by $b_A b_A^T$ in the precision domain'', where the vector $b_A$ was defined in \ref{eqn:ba}. To expand that comment, we use the shorthand $(\cdot)^{*b}$ to denote the operation:
\begin{equation}
\label{eqn:qstarb}
    Q^{*b} :=  \left( Q^{-1} \cdot (b b^T) \right)^{-1}
\end{equation}
then the inter-group allocation matrix sent to $\nu()$ can be written
\begin{equation}
\label{eqn:adash}
  A' = \left(A-BD^{-1}C\right)^{*b_A} 
\end{equation}
or even more briefly 
$$
   A' = \left(A^c\right)^{*b_A} 
$$
where $A^c$ denotes the Schur complement. The matrix $D'$ is computed in analogous fashion. These new matrices will be used to determine the inter-group allocation ratio:
$$
    1/ \nu(A') : 1/ \nu(D').
$$
where $\nu(\cdot)$ is the inverse fitness measure that, as with HRP, is one of the components that can be chosen by the user. 

As with \ref{eqn:topdownbasic}, equation \ref{eqn:topdownschur} implies recursive use and, also in keeping with \cite{DePrado2016BuildingSample} it is anticipated that some variety of seriation will be employed to minimize the typical magnitude of terms in the off-diagonal matrix $B$. The definition of $A'$ in equation \ref{eqn:adashdash} will be relaxed momentarily, but as it stands may look familiar. The numerator is a Schur complement and arises in the expression for conditional covariance.

It is shown in Appendix \ref{sec:examples} that \ref{eqn:topdownschur} with \ref{eqn:adash} and \ref{eqn:adashdash} determining $A'$ and $A''$ can answer the aesthetic concerns pointed out in \ref{sec:critique} and restore lost symmetry. We term this a Hierarchical Minimum Variance (HMV) portfolio or ``Schur complementary'' portfolio if we wish to emphasize the role played by the Schur complement in salvaging off-diagonal information. 

Formal motivation will be provided in Section \ref{sec:linear} but here is a second thought experiment intended to motivate augmenting $A$ and $D$ before they are passed down. Assume that one group of assets represents Europe and the other the U.S. (an unlikely outcome of a mechanical seriation, but it suffices as an example). Perhaps we have decided to allocate $60$\% of our portfolio to U.S. stocks and  $40$\% to Europe. The European sub-allocation is then determined and, we now imagine, we are just about to complete the calculation and sub-allocate the $60$\% amongst U.S. stocks when we are interrupted. At that moment, an oracle informs us of the next period's future returns for the European part of the portfolio. 

How do we react? If we are still free to reallocate within our U.S. portfolio, we will be awfully tempted to use the information that has befallen us. This exercise may seem somewhat irrelevant because one cannot know the future, and certainly there are {\em some} calculations that we can never do in advance using the information revealed. However the conditional covariance {\em is} known in advance. With that seed of doubt planted, we turn to a more formal justification for augmenting $A$ and $D$.

\section{Relationship to the minimum variance portfolio} 
\label{sec:linear}

Some readers might wonder if there needs to be any formal justification for HMV, or for that matter any top-down scheme. It has been the convention to provide none at all, because the benefits are presumably to be demonstrated in numerical simulations, as in \cite{DePrado2016BuildingSample}, or empirical work. 

We observe, however, that HRP and other hierarchical schemes using \ref{eqn:topdownbasic} pay homage to the minimum variance portfolio in the sense that a ``justification'' of splitting and sub-allocating can be obtained down those lines by imagining that $B \approx 0$ (the intent of the seriation). Then HRP as per \cite{DePrado2016BuildingSample} is viewed as a hierarchical way to approximate the minimum variance portfolio (never mind the ``parity'' in the name), and this approximation can be exact if, indeed, off-diagonal covariances are zero and if $\nu()$ is chosen to be portfolio variance.     

If HRP is viewed as approximate hierarchical variance minimization then Schur Complementary Portfolios can be motivated as a graceful cousin that is sometimes {\em exactly} variance minimizing even when $\Sigma$ is arbitrary. Modulo technical considerations (rank and properties of $\Sigma$ and Schur complements of sub-matrices) it is possible to recover \ref{eqn:one} precisely. To do so might serve limited purpose however, if our concern, as with \cite{DePrado2016BuildingSample}, lies primarily with cases where it is unwise or impossible to invert $\Sigma$ at all, or to do so with confidence.     

Instead, we can utilize \ref{eqn:topdownschur} more in the spirit of HRP and we can soften the augmentation of the $A$ and $D$ matrices by introducing parameters that control the extent to which off-block-diagonal covariance information is used. In HRP a split of assets into two groups removes any ability for one group of assets to talk to the other at any further stage of the allocation. Not so in HMV. 

The following recursive representation of the long-short minimum variance portfolio, whose proof we defer, is the key to the connection to the minimum variance portfolio and it is also to be read as a recipe for the construction of new hierarchical schemes:
\begin{equation}
\label{equ:topdown2}
w(\Sigma) \propto \m{ 
            \frac{1}{\nu\left((A^c)^{*b_A}\right)} w\left( \frac{A^c}{b_A b_A^T} \right) \\
          \frac{1}{\nu\left((D^c)^{*b_D}\right)} w\left(\frac{D^c}{b_D b_D^T} \right)
          }                            
\end{equation}
where all terms have been described already. This looks every bit like a recipe for a top-down scheme, as with \ref{eqn:topdownschur}, with $A'$, $A''$ and $D'$, $D''$ taking on the forms we have posited. However \ref{eqn:topdownschur} was merely a suggestion, like \ref{eqn:topdownbasic}, whereas in $\nu()$ is minimum variance portfolio variance then \ref{equ:topdown2} is a mathematical fact about minimum variance portfolios arising from a matrix block inversion identity, provided that $\Sigma$ permits the calculations.  

This is the reason why replication of the awkward minimum variance examples in Section \ref{sec:examples} is possible with HMV but not with HRP. The former therefore allows deviation from minimum variance portfolios to be a more conscious choice, which is the next task we turn our attention to, and it suggests families of recursive allocation schemes where portfolio variance $\nu()$ can be exchanged for some other fitness measure, seriation can be used to good effect, and a choice of terminal portfolio $w_{term}$ is also up for grabs.   

To emphasize, the literal use of \ref{equ:topdown2} is not advocated and nor, depending on inputs, is this possible to compute in generality - especially if the terminal portfolio constructor $w_{term}$ is an off-the-shelf optimizer. As a practical matter this depends on the assumptions made by portfolio allocation packages employed, or the explicit checks they perform (some, but not all, demand positive definite matrices, for example).

\begin{table}[]
    \centering
    \begin{tabular}{|p{4cm}|p{2cm}|p{5cm}|}
    \hline 
         Allocation    &  HRP & Schur    \\
     \hline 
     Inter-group     &  $A$ or $diag(A)$ &  $(A^c(\gamma)^{-1} \cdot b_A  b_A^T)^{-1}$                where we set $A^c(\gamma)=A-\gamma B D^{-1} C$ and $b_A(\lambda) = \one - \lambda B D^{-1} \one$. 
           \\
     \hline 
     Intra-group  &  $A$            &  $ \left(A-\gamma B D^{-1} C\right)/\left(b_A b_A^T\right)$ element-wise division \\
     \hline
    \end{tabular}
    \caption{Comparison of covariance matrices used to allocate capital between and within a group of securities in two divide-and-conquer schemes: HRP as proposed in \cite{DePrado2016BuildingSample} and Schur Hierarchical Portfolios as proposed herein.}
    \label{tab:schur_matrices}
\end{table}

Instead, a compromise between hierarchical and optimization is established by introduction of one or more parameters. In the remainder we will take $\gamma \in [0,1]$ and modify both the Schur complement
\begin{equation}
\label{eqn:Agamma}
  A^c(\gamma) = A - \gamma BD^{-1} C
\end{equation}
and also the $b$'s:
\begin{equation}
\label{eqn:bgamma}
     b_A(\gamma) := \one -  \gamma B D^{-1} \one 
\end{equation}
and similarly for $D^c$ and $b_D$. These matrices, and their roles, are also summarized in Table \ref{tab:schur_matrices}. If $\gamma=0$ we recover the usual hierarchical family of methods. On the other hand if $\gamma=1$ we can recover the minimum variance portfolio (caveats already noted), as noted in Table \ref{tab:extremal}.\footnote{In the accompanying code a method of determining $A^c(\gamma)$ is provided that is adaptive. We first find the largest possible $\gamma < 1$ subject to the constraint that $A^c$ is positive definite. This is then scaled by a parameter, which is how the $\gamma$ in the code should be interpreted.}

\begin{table}[]
    \centering
    \begin{tabular}{|l|p{5cm}|l|}
    \hline 
      Method   &  Methodology & Gamma \\
      \hline 
     Hierarchical risk parity & Recursive allocation, typically with ``diagonal'' allocation. & $\gamma=0$ \\ 
     \hline 
     Minimum variance & One-off global constrained minimization of portfolio variance. & $\gamma=1$ \\
     \hline 
    \end{tabular}
    \caption{Extremal points on the Schur Hierarchical Minimum Variance (HMV) continuum spanning both Hierarchical Risk Parity and explicit optimization (rank permitting). Corresponding $\gamma$ for HMV used in \ref{eqn:Agamma} and \ref{eqn:bgamma} are shown. HMV can use seriation to reorder assets, an important ingredient in HRP.}
    \label{tab:extremal}
\end{table}

\section{Simulation study in a domain where HRP is preferred}
\label{sec:numerical1}

Code is provide in \cite{cottonprecise} following a similar path to the experiment in \cite{DePrado2016BuildingSample}. Comparisons were arranged between otherwise identical HMV methodologies varying only in the choice of $\gamma$ that appears in \ref{eqn:Agamma} and \ref{eqn:bgamma}. This HMV family nests HRP when $\gamma=0$ and thus permits a direct comparison. 

As with that motivating paper we do not use aggressive shrinkage of covariance. This choice may be viewed as controversial by those who consider the concentrated portfolios in \cite{DePrado2016BuildingSample} to be a poor benchmark. As a compromise we use ``weak'' shrinkage defined in Appendix \ref{sec:weak}. This serves to steer the optimized portfolio away from cases where it is clearly too opinionated, as evidenced by pronounced long-short positions, as also discussed in \ref{sec:weak}. We also use this weakening to define the portfolio fitness $\nu()$, namely as the variance of the minimum variance portfolio when this weak shrinkage is applied. 

\begin{figure}
    \centering
    \includegraphics[scale=0.5]{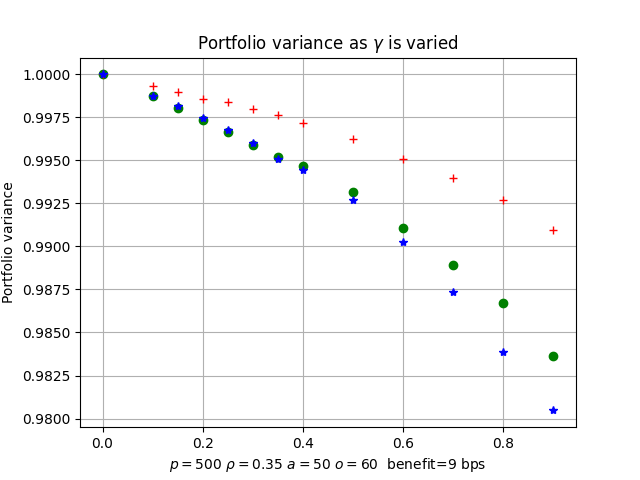}
    \caption{Three different examples of relative portfolio variance as the $\gamma$ parameter (x-axis) is varied from $\gamma=0$ (traditional top-down allocation) to $\gamma=1$ (using as much information from the Schur complement as allowed). Numbers are normalized by the portfolio variance for $\gamma=0$. In this example $p=500$ assets are used with a true covariance matrix assumed to be the empirical covariance of $a=50$ samples from a symmetric model with constant off-diagonal correlation $\rho=0.35$. Then $o=60$ observations are used to estimate $\Sigma$, the covariance matrix used for top-down allocation per \ref{eqn:topdownschur}. The approximate economic benefit is on the order of $10$bps of additional return per year, with these settings.}
    \label{fig:varying_gamma}
\end{figure}

In the particular experiment generating Figure \ref{fig:varying_gamma} the following steps are taken:

\begin{enumerate}
    \item An approximately constant off-diagonal covariance matrix dimension $p=500$ is generated, with off-diagonal correlations averaging close to $\rho=0.35$. See \verb|rand_symm_cov| in \cite{cottonprecise} for preicse details.   
    \item We set $\Sigma_{true}$ to be the empirical covariance of $a=150$ samples from this anchor matrix. 
    \item We then generate $o=60$ samples from $\Sigma_{true}$ and the covariance estimate from these samples is used as input to the Schur Hierarchical Portfolio $w$ for various values of $\gamma$. We use $\nu()$ 
    \item Out of sample variance $w^T \Sigma_{true}w$ is measured.
\end{enumerate}

Results for three different randomly generated true covariance matrices $\Sigma_{true}$ are overlaid on the plot. In each case top down allocation with \ref{eqn:topdownschur} was computed terminating when the portfolio size was $5$, at which point the minimum variance portfolio is used to sub-divide, also with weak shrinkage (thus approach in Section \ref{sec:weak} is used once for the portfolio variance and then again for the terminal portfolio choice $w_{term}$).

As can been seen from Figure \ref{fig:varying_gamma}, the inclusion of information using the Schur complement (i.e. taking $\gamma > 0$) steadily reduces portfolio variance. The difference is material for institutional investors, and corresponds to a roughly equivalent increase in portfolio return on the order of $10$ basis points (bps). For perspective, those gains carry economic value on the order of tens of millions per year, however, for funds that are roughly on the same scale as, say, the larger bespoke ETFs. 

In this comparison it is worth noting that all the choices of $\gamma$ are superior to a naive optimization, in keeping with \cite{DePrado2016BuildingSample} and also superior to equal-weighting of assets. The numbers above speak only of the incremental improvement that applies to a specific choice of top-down construction in the tradition established by \cite{DePrado2016BuildingSample}.

It is difficult to anticipate the choices of shrinkage, or other modifications to $\Sigma$ that the reader might devise, and the impact on the outcome of applying \ref{eqn:topdownschur} recursively from that starting point. Again, code is provided for that purpose \cite{cottonprecise} but given all the possible permutations that are possible, the cautious statement is that a choice of $\gamma>0$ seems almost invariably to help in the set of those idealized simulation settings where top-down HRP beats optimization (or simple benchmarks like ignoring all off-diagonal entries). 

The cautious conclusion is that investors with a strong conviction in top-down allocation should give serious consideration to \ref{eqn:topdownschur} over \ref{eqn:topdownbasic}, and not only because of the aesthetic advantage of the former.

We do not anticipate that all readers will be compelled by numerical experiments to abandon traditional optimization-base approaches, any more than those in \cite{DePrado2016BuildingSample} would convince them to, for two primary reasons. 

The first is that optimization performed on naked covariance estimates is well appreciated to be sub-optimal, per the discussion in Section \ref{sec:literature}. 

The second is that any such numerical experiment presupposes an environment where certain assumptions are made about the extent to which we can discern the true covariance matrix (not to mention the true returns), and these assumptions may not reflect reality. 

One crucial design decision in these experiments, which the reader is welcome to modify, is the nature and extent of knowledge of the true $\Sigma$ and the generating process for the same. In the code, as noted, the former is parameterized by a number of samples but it should be understood that this is intended as an {\em equivalent} number of samples given that there exists plenty of exogenous data that can inform the estimation of asset volatility and covariance. 

Thus we do not claim here that Schur Hierarchical Portfolios represent a strictly better way to assign weights in all contexts, and nor - returning to our thought experiment - should anyone believe that. Clearly the choice of best method in the low information extreme cannot be the same as the high. 

The relevant question is whether HMV, as a bridge between direct optimization and heuristic splitting schemes such as HRP, is likely to provide a useful, practical, middle ground.

\section{Conclusion}
Despite seminal work in \cite{Markowitz1952PortfolioFinance} and subsequently \cite{merton1969lifetime} the use of optimization for real world portfolio allocation is limited. In a recent paper John H. Cochrane went so far as to state that the textbook solutions had ``almost no impact on portfolio practice''. Instead, the author describes a typical institutional investment pattern \cite{cochrane2022portfolios}. 

\begin{quote}
Roughly, they start with intense attention to buckets, defined by names rather than betas and correlations: debt and equity; subcategories mostly denoted by industry and other nonbeta characteristics, but now also growth; value
and other academic styles; domestic, foreign; “alternatives,” often just an alternative organizational forms which repackage the same securities; real estate, private equity, venture capital, etc. Having decided these buckets down to the last percentage point, these investors
allocate each bucket to separate funds or managers, evaluated by rather short-term returns
relative to a benchmark, and frequently replaced based on short-term results.
\end{quote}

This type of hierarchical decision making is the norm and there is little formal insight to guide its use, or ameliorate the obvious shortcomings we have discussed in Section \ref{sec:critique} including the unfortunate discarding of information. Even the literature discussed in Section \ref{sec:literature} is numerically or empirically motivated, and has grown around \cite{DePrado2016BuildingSample} largely in detachment from the rest of theory, at least as far as the overall portfolio is concerned (if not the components, such as covariance estimation or seriation that might be utilized).   

Given this void, it is hoped that the new family of hierarchical portfolio construction techniques we have provided, and in particular the idea of augmenting covariance matrices, can provide a relatively non-invasive way to augment top-down allocation schemes, permitting the cautious introduction of off-diagonal covariance information. We have exhibited Hierarchical Minimum Variance (HMV) nesting the well-known Hierarchical Risk Parity method. 

The use of Schur complement-inspired altered sub-covariance matrices used in inter-group allocation and sub-portfolio determination is not limited to cases where an approximation to the minimum variance portfolios is desired, but in that case it does provide an elegant connection back to Modern Portfolio Theory. In that sense, hierarchical methods are no longer their own island.

\section{Appendix A: Split representation of a minimum variance portflio} 
\label{sec:split}

Here we unravel the identity \ref{equ:topdown2} to arrive at the suggestions \ref{eqn:adashdash} and \ref{eqn:adash}. Take \ref{eqn:one} and apply a block inversion identity:
\begin{equation*}
  \m{ A & B \\ C & D }^{-1} = \m{ (A-B D^{-1} C)^{-1} & 0 \\ 0 & (D - C A^{-1}B)^{-1} } \m{ 1 & -B D^{-1} \\ - C A^{-1} & 1  }
\end{equation*}
Denoting the Schur complement $A^c = A-BD^{-1}C$ and analogously for $D^c$ we see immediately that \ref{eqn:one} can be reformulated: 
\begin{equation}
\label{eqn:inverses}
 w \propto \Sigma^{-1} \one = \m{ (A^c)^{-1} \left( \one - B D^{-1} \one \right) \\
                               (D^c)^{-1} \left( \one - A C^{-1} \one \right) }
    = \m{ (A^c)^{-1} b_A \\
                               (D^c)^{-1} b_D }
\end{equation}
We are almost back to \ref{equ:topdown2} already, but must recognize the solution of the linear systems on the right hand side as scaled minimum variance portfolios. To that end, consider the long-short portfolio with unequal constraints:
\begin{equation}
   w* = \arg \min_w  w^T Q w \ s.t.  w^T b = 1 
\end{equation}
with solution
\begin{equation}
\label{eqn:wqb}
    w* = w(Q,b) = \frac{Q^{-1}b}{b^T Q^{-1} b}
\end{equation}
The variance of said portfolio is
\begin{equation}
\label{eqn:qb}
   \nu(Q,b) = w(Q,b)^T Q w(Q,b) = \frac{1}{b^T Q^{-1} b}
\end{equation}
so rearranging \ref{eqn:qb} and \ref{eqn:wqb}, we have a useful insight: the solution to any symmetric linear system $Qx=b$ carries a financial interpretation:
\begin{equation}
\label{eqn:qinv}
    Q^{-1} b = \frac{1}{\nu(Q,b)} w(Q,b)
\end{equation}
Now using this to interpret \ref{eqn:inverses} we arrive at \ref{equ:topdown2}. 

The accompanying code provides an alternative \cite{cottonprecise}. It is also possible to collapse Equation \ref{eqn:inverses} a different way, viz: 
$$
   (A^c)^{-1} \left( \one - B D^{-1} \one \right) =  \left(\tilde{A}\right)^{-1} \one 
$$
so long as a $\tilde{A}$ can be constructed with this property. And from that point we follow a similar path, interpreting the right hand side as the scaled solution of a slightly different minimum variance problem.

\section{Appendix B: Explanation of the Schur augmentation by example}

\label{sec:examples}

The smallest non-trivial example suffices to illustrate how \ref{eqn:adashdash} and \ref{eqn:adash} can reproduce the minimum variance portfolio whereas Hierarchical Risk Parity does not. Take:
$$
\Sigma = \m{1 & \rho & \rho \\ \rho & 1 & \rho \\ \rho & \rho & 1}
$$
The desired portfolio is the symmetric one, clearly. However if we apply a bisection approach such as HRP we must break the symmetry. Without loss of generality this splitting into groups is $\{1,2\},\{3\}$ and, consequently, the covariance splits per \ref{eqn:split} into 
$$
    A = \m{ 1 & \rho \\ \rho & 1 }, B = \m{\rho \\ \rho}, C = \m{\rho, \rho}, D = \m{1}
$$
Most hierarchical methods will use a sub-allocation within $\{1,2\}$ that will allocate evenly amongst the two assets (since $A$ is symmetric). Then every dollar allocated towards this part of the portfolio, as compared with the third asset, incurs variance $v_A =  \frac{1+\rho}{2}$. A dollar invested in the third asset incurs unit variance, naturally. Thus a seemingly reasonable hierarchical portfolio allocation assigning capital inversely proportional to variance will lead to an inter-group portfolio allocation:
$$
      \frac{2}{3 + \rho} :  \frac{1 + \rho}{3+\rho}
$$
Then, splitting the allocation to $\{1,2\}$ in half we have 
$$
  w = \frac{1}{3+\rho} \m{ 1 \\ 1 \\ 1 + \rho }
$$
So, despite the reasonableness of this methodology, it evidently over-allocates to asset $3$ when $\rho>0$ and under-allocates when $\rho<0$. 

This inelegance is fixed in the new approach. We can see explicitly how this is accomplished, and how it relates to the matrix inversion identity motivating \ref{eqn:topdownschur}, \ref{eqn:adashdash} and \ref{eqn:adash}. We compute the Schur complement inverse:
\begin{eqnarray*}
 (A^c)^{-1} & = &  
    \frac{1}{\phi(\rho)} \m{ 1+\rho & -\rho \\ -\rho & 1+\rho }
\end{eqnarray*}
where $\phi(\rho) := 1 + \rho - 2\rho^2$. Completing the rest of the algebra the identity
\begin{eqnarray*}
   \Sigma^{-1} & = & \m{A & B \\
                          C & D}^{-1} =  \m{ (A^c)^{-1} & 0 \\
                             0 & (D^c)^{-1} } \m{ 1 & -BD^{-1} \\
                                                 -CA^{-1} &  1 } 
\end{eqnarray*}
manifests in this example as
\begin{eqnarray*}
  \m{ 1 & \rho & \rho \\
      \rho & 1 & \rho \\
      \rho & \rho & 1 }^{-1} & = & \frac{1+\rho}{\phi}  \m{ 1 & -\frac{\rho}{1+\rho} & 0 \\
                                          -\frac{\rho}{1+\rho} & 1 & 0 \\
                                          0 & 0 & 1  } 
                                          \m{  1 & 0 & -\rho \\
                                               0 & 1 & -\rho \\
                                               -\frac{\rho}{1+\rho} & -\frac{\rho}{1+\rho} & 1} 
\end{eqnarray*}
The right hand side has broken symmetry but we can check that weights of the minimum variance portfolio are equal:
\begin{eqnarray*}
\label{eqn:proportion}
  w \propto \Sigma^{-1} \one & \propto & \m{ 1 & -\frac{\rho}{1+\rho} & 0 \\
                                          -\frac{\rho}{1+\rho} & 1  & 0 \\
                                          0 & 0 & 1 }  \m{  1 & 0 & -\rho \\
                                               0 & 1 & -\rho \\
                                               -\frac{\rho}{1+\rho} & -\frac{\rho}{1+\rho} & 1} \m{1 \\ 1 \\ 1} 
                         =  \frac{1-\rho}{1+\rho}\m{1 \\ 1 \\ 1}
\end{eqnarray*}
as they should be.

\section{Appendix C: Weak covariance shrinkage}
\label{sec:weak}

\begin{figure}
    \centering
    \includegraphics[scale=0.7]{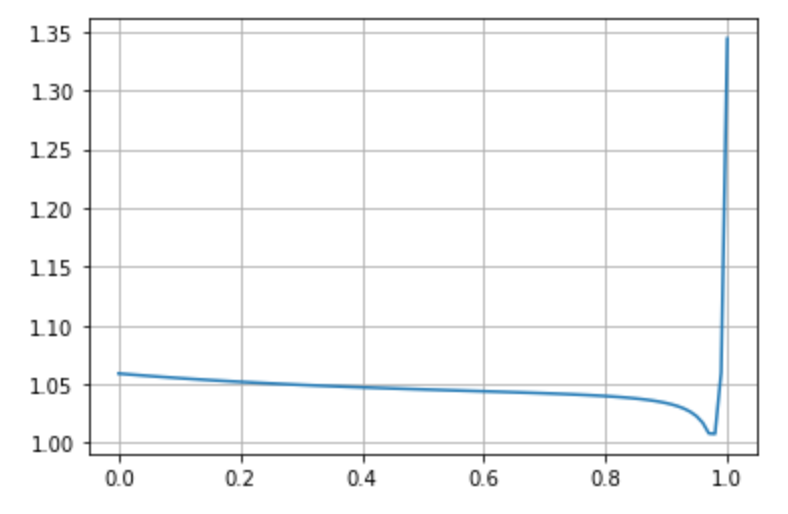}
    \caption{Defining ``weak'' adaptive shrinkage. The plot shows the portfolio variance of the long-only portfolio that is achieved by first multiplying off-diagonal entries and then redistributing the mass. The minimum in this case occurs for a shrinkage of $0.97$.  }
    \label{fig:weak}
\end{figure}

Shrinkage was used in some of the simulation studies. As the method is non-standard we briefly motivate it. The intent was to provide a minimalist augmentation only, but not allow examples that are likely to overly penalize optimization. Consider:
$$
  \Sigma = \m{ 1.09948514 &  -1.02926114 &   0.22402055 &   0.10727343 \\
       -1.02926114 &   2.54302628 &   1.05338531 &  -0.12481515 \\
        0.22402055 &   1.05338531 &   1.79162765 &  -0.78962956 \\
        0.10727343 &  -0.12481515 &  -0.78962956 &   0.86316527 
}
$$
The minimum variance portfolio is
$$
    w = \frac{\Sigma^{-1} \one}{\one^T \Sigma^{-1} \one} = \m{-9.008 \\ -6.871 \\   8.749 \\   8.130 }
$$
This solution is highly unstable as we can see by multiplying off-diagonal entries by $\xi=0.97$ whereupon the minimum variance portfolio is:
$$
   w = \m {0.0674 \\ -0.0068 \\  0.3658 \\  0.5735}
$$
The choice $\xi=0.97$ is suggested by Figure \ref{fig:weak}. The value is the exact amount of shrinkage required to minimize the variance of a portfolio {\em after} short positions have been nullified and the corresponding positive mass redistributed to all assets (the variance is judged by the original $\Sigma$, not the shrunk matrix). This choice of shrinkage usually leaves only small short positions. The approach is different to \cite{DeMiguel2009ANorms} where the norm of weights are constrained, but shares a similar goal. 

\bibliographystyle{apalike}
\bibliography{refs}
\end{document}